\documentclass[sigconf, authorversion]{acmart}
\AtBeginDocument{%
  \providecommand\BibTeX{{%
    \normalfont B\kern-0.5em{\scshape i\kern-0.25em b}\kern-0.8em\TeX}}}

\setcopyright{acmlicensed}
\copyrightyear{2024}
\acmYear{2024}
\acmDOI{3652988.3673949}

\usepackage[normalem]{ulem}
\useunder{\uline}{\ul}{}
\usepackage{hyperref}
\usepackage{tabularx}
\usepackage{array}
\usepackage{natbib} 
\usepackage{float}

\acmConference[Preprint]{}{June 2024}{}
\acmBooktitle{Preprint} 
\acmISBN{979-8-4007-0625-7/24/09}

\newcommand{\repourl}{https://github.com/arminarj/empathic-grounding}

\makeatletter
  \if@ACM@anonymous
    \renewcommand{\repourl}{https://anonymous.4open.science/r/empathic-grounding-3383/}
  \fi
\makeatother

\begin{document}

\title[Empathic Grounding]{Empathic Grounding: Explorations using Multimodal Interaction and Large Language Models with Conversational Agents}

\author{Mehdi Arjmand}
\affiliation{%
  \institution{Northeastern University}
  \country{Boston, MA}}
\email{arjmand.me@northeastern.edu}

\author{Farnaz Nouraei}
\affiliation{%
  \institution{Northeastern University}
  \country{Boston, MA}}
\email{nouraei.f@northeastern.edu}

\author{Ian Steenstra}
\affiliation{%
  \institution{Northeastern University}
  \country{Boston, MA}}
\email{steenstra.i@northeastern.edu}

\author{Timothy Bickmore}
\affiliation{%
  \institution{Northeastern University}
  \country{Boston, MA}}
\email{t.bickmore@northeastern.edu}

\renewcommand{\shortauthors}{Arjmand et al.}

\newcommand{\armin}[1]{\textcolor{magenta}{~Armin:~#1}}
\newcommand{\farnaz}[1]{\textcolor{red}{~Farnaz:~#1}}
\newcommand{\ian}[1]{\textcolor{green}{~Ian:~#1}}
\newcommand{\tim}[1]{\textcolor{orange}{~Tim:~#1}}

\begin{abstract}

We introduce the concept of ``empathic grounding'' in conversational agents as an extension of Clark's conceptualization of grounding in conversation in which the grounding criterion includes listener empathy for the speaker's affective state. Empathic grounding is generally required whenever the speaker's emotions are foregrounded and can make the grounding process more efficient and reliable by communicating both propositional and affective understanding.
Both speaker expressions of affect and listener empathic grounding can be multimodal, including facial expressions and other
nonverbal displays. Thus, models of empathic grounding for embodied agents should be multimodal to facilitate natural and efficient
communication.
We describe a multimodal model that takes as input user speech and facial expression to generate multimodal grounding moves for a listening agent using a large language model.
We also describe a testbed to evaluate approaches to empathic grounding, in which a humanoid robot interviews a user about a past episode of pain and then has the user rate their perception of the robot's empathy.
We compare our proposed model to one that only generates non-affective grounding cues in a between-subjects experiment. Findings demonstrate that empathic grounding increases user perceptions of empathy, understanding, emotional intelligence, and trust. Our work highlights the role of emotion awareness and multimodality in generating appropriate grounding moves for conversational agents.

\end{abstract}





\begin{teaserfigure}
    \centering
    \includegraphics[width=\textwidth]{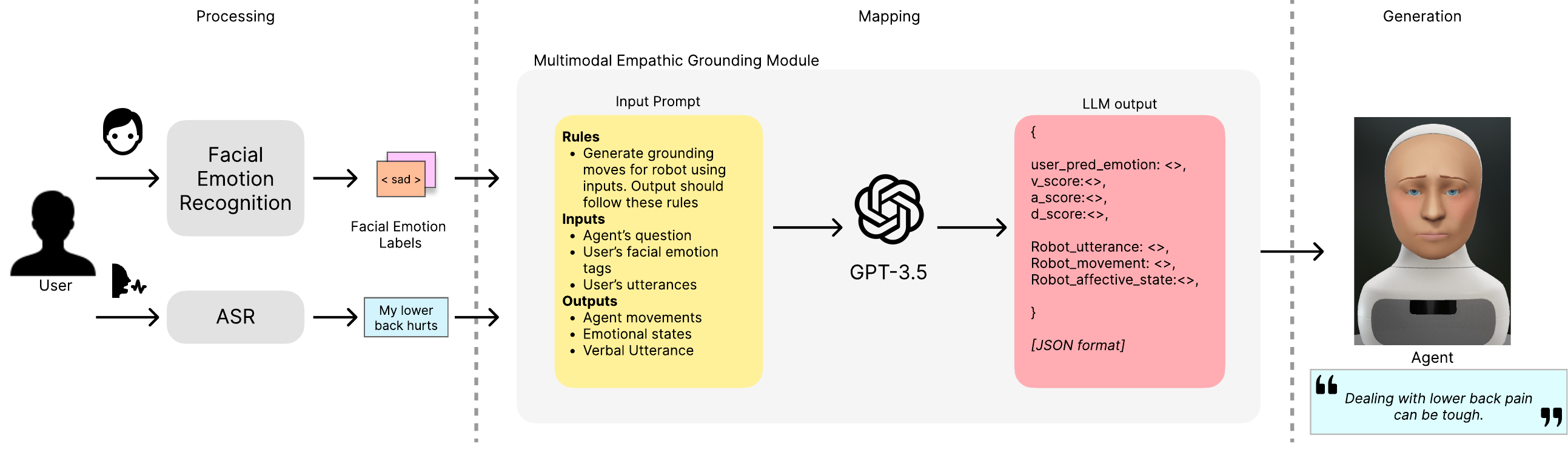} 
    \caption{Computational Model of Multimodal Empathic Grounding }
    \label{fig:teaser}
\end{teaserfigure}

\received{April 2024}

\maketitle

\section{Introduction}

Empathy is the understanding and/or feeling of the affective experiences of another person \cite{Wispe1986}. At least three kinds of empathy have been identified in the literature: \textit{cognitive empathy}, which is the perception and understanding of another's affective states, \textit{affective empathy}, which is the experiential alignment with another's affective states (i.e., actually feeling what the other is feeling), and \textit{narrative empathy}, which is empathy for others in related narratives and experiences \cite{Keen2010}.
Empathy is hard-wired into the brain; the mere act of observing affective facial displays triggers neural responses in the observer's own emotional circuitry, including activation of premotor areas related to the observer's own facial displays mirroring those of the observed \cite{Carr2003}.
A closely-related concept to empathy is \textit{sympathy}, which is the awareness of another's negative affective state as something to be alleviated \cite{Wispe1986},
often by the listener conveying their feeling about the speaker's affective state, such as concern or compassion.
However, most recent work in psychology and the helping professions conflate sympathy and empathy, with the latter term having dominant use.

The literature on empathy is voluminous, but can be roughly categorized into two levels of analysis. At the \textbf{macro level}, empathic communication encompasses intentional, overt
conveyance of understanding of another's emotional state to satisfy instrumental goals of the listener, and can span many turns of dialog.
Examples include acknowledgment of patient distress by
healthcare providers, ''reflections'' in counseling (e.g., motivational interviewing \cite{Miller2023}), and active listening.
In medicine, physician empathy for patients has been associated with greater patient satisfaction, better health outcomes, and fewer malpractice suits \cite{Lorie2017},
and is important enough that training in empathic communication has been mandated as a learning objective for medical school education \cite{Riess2012}.

At the \textbf{micro level}, empathy conveys understanding of some aspect of the speaker's current or related affective state that the they just communicated.
This process is largely unconscious and the empathic display can be very brief, consisting of as little as a transitory facial display by the listener. In this role, empathic cues play a role in \textit{grounding}: the process interlocutors use to incrementally update their mutual knowledge during  conversation
\cite{Clark1991}.
For an utterance to be understood, the interlocutors
must arrive at a situation in which they mutually believe that
the listener understood what the speaker meant to a criterion sufficient for current purpose, something referred to as the \textit{grounding criterion}. 
For utterances in which the speaker's current or related affective state is communicated or implied, the grounding criterion may include listener empathy. 

Clark and Brennan identified five types of evidence that human listeners can provide to demonstrate that they understood a speaker's last utterance (\autoref{tab:grounding}).
Listener empathy can serve, in whole or in part, as grounding cues for levels 2-5. 

\begin{table*}
\resizebox{\textwidth}{!}{
\begin{tabular}{|c|l|l|}
\hline
 \textbf{Level} & \textbf{Evidence}  & \textbf{Description} \\ \hline
1 & Continued attention  & H shows he/she is continuing to attend and therefore remains satisfied with S's presentation.    \\ \hline
2 & Initiation of the relevant next contribution  & H starts in on the next relevant contribution.  \\ \hline
3 & Acknowledgment  & H nods or says ''uh huh'' (backchannel).    \\ \hline
4 & Demonstration  & H demonstrates all or part of what he/she has understood S to mean.    \\ \hline
5 & Display  & H displays verbatim all or part of S's presentation.    \\ \hline
\end{tabular}}
\caption{Evidence of Understanding in Grounding, ranked from weakest to strongest, for Speaker S and Hearer H. Based on \cite{Clark1991}}.
\label{tab:grounding}
\end{table*}

Clark extended the Gricean maxims of Quantity and Manner \cite{Grice} to define the principle of least collaborative effort, where
interlocutors attempt to minimize the joint effort they invest in making a conversation work. 
Since empathy or sympathy can be displayed nonverbally (e.g., via facial display), these cues facilitate grounding by allowing grounding evidence to be
distributed across channels simultaneously. 
For example, brief affect displays can be used as acknowledgments (``backchannel'' cues) that do not interrupt the speaker, but convey more information
than affectively-neutral cues such as headnods.
Smiles have been studied as backchannel cues \cite{Bertrand2007}, \cite{Bilalpur2024}, as have nonverbal displays of positive and negative affect \cite{Shahverdi2023},  but we argue that all affective displays can not only play this role, but also provide more grounding evidence than affectively neutral backchannels, regardless of modality.

We introduce the concept of \textbf{empathic grounding} as an extension of Clark's conceptualization of grounding in which the grounding criterion includes listener empathy for the speaker's affective state.
This encompasses both micro-empathy and affective backchannels, but also macro-empathy and sympathy.
Empathic grounding is not always required, since affectively neutral information exchanges can be grounded without affect.
However, it facilitates communication and the grounding process whenever the speaker's emotions are foregrounded or when the speaker is relating affectively-charged narratives or information. 
In addition, empathic grounding is essential for macro-empathy.

Relative to Clark and Brennan's model (\autoref{tab:grounding}), empathic communication comprises demonstration or display grounding cues, by definition.
We note that brief empathic facial displays, with no verbal accompaniment, fit between acknowledgments and demonstrations in strength, since they convey valenced information--beyond simple acknowledgment--that shows some level of understanding. We agree with \cite{Allwood1992} that demonstrations typically provide stronger information than displays, at least for empathic grounding, since demonstrations require a level of interpretation that displays do not. 

We differentiate empathic grounding from Jung's ``Affective Grounding'' in HRI \cite{Jung}. Jung's work focused on emotion communication and regulation among interlocutors, but as a separate track from non-emotional communication and grounding processes.
Here, we argue that propositional and affective communication are entagled and support each other in empathic grounding. In particular, we are interested in the role of affective communication in enhancing propositional communication and grounding processes.

In this paper, we present a computational model of multimodal empathic grounding for an embodied conversational agent that uses a large language model (LLM) to generate grounding responses for user utterances. Prior studies indicate that LLMs may be good at generating empathic language, with one study demonstrating that
LLM-generated responses to medical questions from laypersons were rated higher on empathy than those generated by physicians \cite{Ayers2023}.
Our model goes beyond text, incorporating information from user speech and nonverbal displays of affect, and produces brief agent utterances of empathy and sympathy accompanied by appropriate affective and interactional nonverbal behavior.

We also present a testbed for evaluating empathic grounding models in the domain of interviewing participants about a prior painful experience, that isolates the effects of grounding responses for comparison.

We report results from a between subjects experiment in which our empathic grounding model (EMPATHIC GROUNDING) is compared against a baseline control model (BACKCHANNEL) that only generates affectively-neutral backchannels. 

We hypothesize that participants relating their pain experience in the EMPATHIC GROUNDING condition, compared to participants in the BACKCHANNEL condition, will rate the agent higher on
(\textbf{H1}) perception of being understood, (\textbf{H2}) empathy and empathic listening, and (\textbf{H3}) trust. In addition, we hypothesize that
(\textbf{H4}) recognition and incorporation of user affect from multimodal cues will lead to better empathic grounding.


\begin{figure}[tp]
    \centering
    \includegraphics[width=\columnwidth]{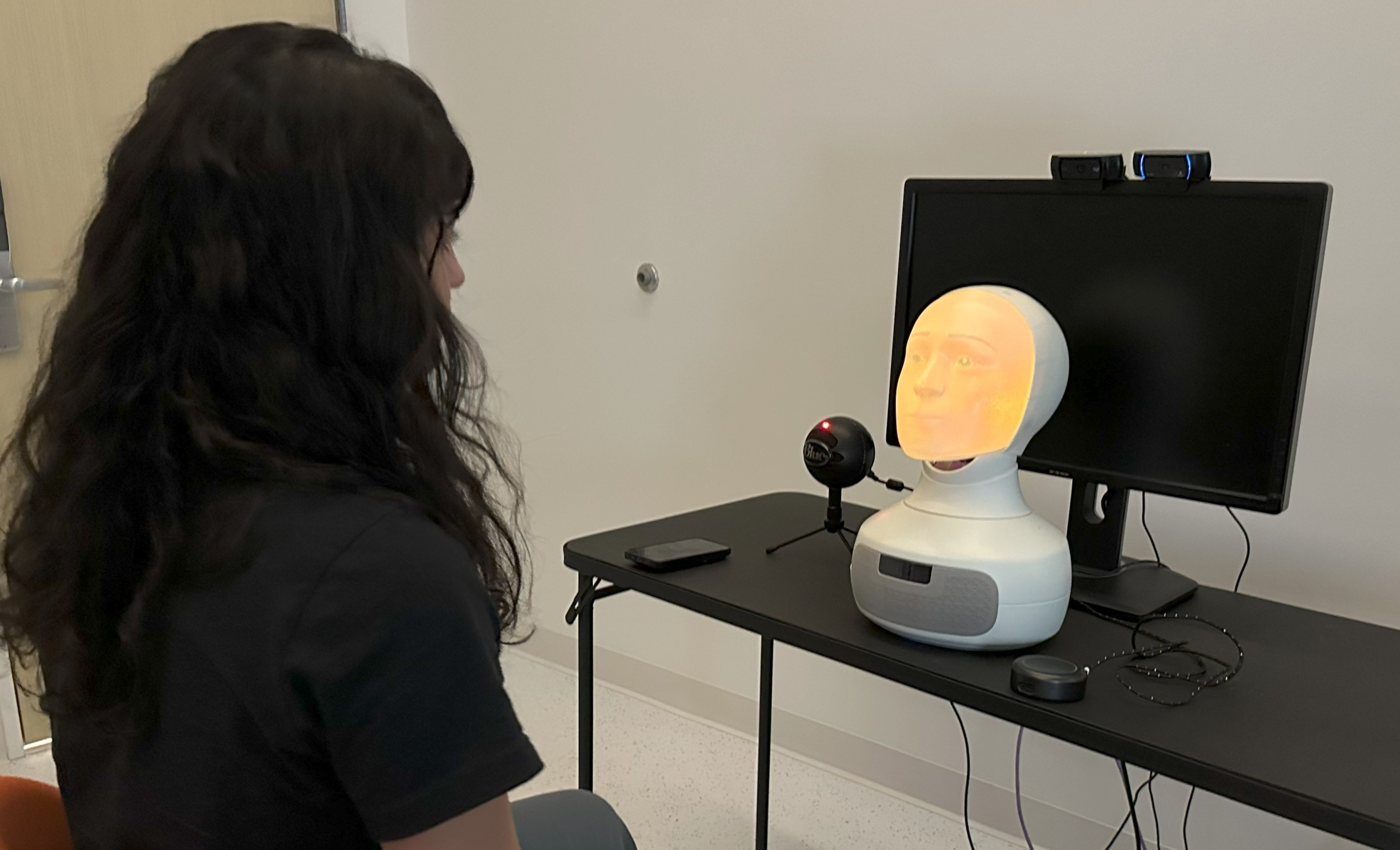}
    \caption{User interacting with the robot testbed for the empathic grounding model}
    \label{fig:interaction}
\end{figure}

\section{Related Work}

\subsection{Grounding in Human-Agent Interactions}
Previous work on conversational agents has seen long-standing interest in the notion of grounding \cite{Clark1991}. Task-based dialog systems require feedback from users to understand the result of their actions \cite{wiener2019cybernetics}. Similarly, dialog systems that provide feedback to users to signal understanding are able to create more responsive and efficient interactions \cite{ward1996using}. In this realm, Traum's computational model of grounding \cite{traum1994computational} provides grounding acts, such as acknowledgments and repairs, to enable grounding in multi-turn human-agent dialog. Building on this model, \cite{nakano2003towards} proposed a conversational agent design that maps the user's verbal and non-verbal behavior to the agent's dialog state to perform grounding acts. However, such a system can only fulfill the propositional aspect of dialog through grounding acts, as both the input and output of the system are unaware of emotion cues.

Among forms of grounding moves, backchanneling has been widely studied in virtual agents. Most work on listener backchannel generation has used rules based on prosodic and linguistic cues of the speaker \cite{Ward2000, lala2017attentive}. The appropriate generation of backchannels by listening agents has been shown to be effective in establishing rapport and increasing engagement with users \cite{gratch2006virtual, huang2011virtual, heylen2007multimodal}. For example, \citeauthor{gratch2007creating} \cite{gratch2007creating} developed a virtual agent that expressed listening and backchanneling behaviors and found that contingent backchanneling with respect to user utterances can effectively build rapport compared to non-contingent feedback \cite{gratch2007creating}. Compared to their approach to providing ``envelope feedback'' and following prior research showing the needs of speakers for relevant listening behaviors in particularly emotional contexts \cite{bavelas2000listeners}, we explore whether more contentful grounding moves, including semantically- and affectively-relevant feedback can establish a higher level of trust and empathy with users. Further, while most prior work has focused on generating appropriate non-verbal backchanneling behaviors, less attention has been given to producing contingent verbal utterances alongside non-verbal cues \cite{heylen2007multimodal}. Our work bridges this gap by introducing an LLM grounding move generation module that takes multimodal user input and generates multimodal agent output in real-time. 

Another thread of research has focused on multimodal emotion processing and delivering feedback to users via empathic listening agents \cite{buschmeier2014elicit, schroder2011building, devault2014simsensei}. For example, \cite{devault2014simsensei} proposed a framework to incorporate users' multimodal input and provide listening feedback, including verbal and nonverbal backchannels. The authors found that the system was able to build rapport with users in a mental health screening context, and led to more self-disclosure. However, due to limitations in speech recognition, meaningful grounding moves were sporadically generated in practice \cite{10.1145/3477322.3477335}, and empathic utterances were limited. In this work, we propose a novel multimodal empathic grounding component that incorporates the user's verbal and non-verbal inputs and generates appropriate empathic grounding behaviors using multiple modalities, including utterances, emotional states, and movements.

\subsection{Multimodal Human-Agent Interaction}
Human face-to-face conversation is highly multimodal, with facial displays, hand gesture, and proxemics playing important roles in structuring interactions and conveying information beyond speech.
Thus, previous research has attempted to use multimodal interaction in conversational agents, including social robots and virtual agents \cite{SIABook1, 10.1145/3563659}. 

Several ways of modeling multimodal conversational behavior have been used, including 
 theory and rule-based \cite{cassell2001beat},  
 and machine learning and corpora-based approaches \cite{lee2006nonverbal}.

Our model of multimodal empathic grounding can be characterized by the computational framework for multimodal interactions in \cite{10.1145/3563659.3563664}, which outlines three primary components: processing, mapping, and generating multimodal signals. Processing involves signal sensing, recognition, and interpretation, while mapping focuses on planning and rule application and generating entails producing output signals. In our model, we perform the processing component explicitly, while mapping and generating are performed using an LLM. We describe the details of our model architecture in \autoref{sec:empathic}.

\subsection{LLMs for Social Agents}

Pre-trained large language models (LLMs) such as GPT-based models \cite{brown2020language, achiam2023gpt}, and LlaMA \cite{touvron2023llama} have revolutionized natural language processing, advancing tasks related to understanding and generating language. Specifically, the perception of empathy and emotional intelligence in LLMs has been of rising interest among researchers in social dialogue systems. For instance, studies have shown that LLM-generated responses to social media posts describing common life experiences are consistently perceived as more empathic than human-written ones \cite{lee2024large}. \citeauthor{cuadra2024illusion} demonstrate a systematic analysis of LLMs display and perception of empathy \cite{cuadra2024illusion}. Moreover, multiple studies demonstrated the capabilities and limitations of LLMs in text-based emotion recognition tasks \cite{tak2023gpt, broekens2023fine, zhan2023evaluating}. Recently, conversational agents have used LLMs to improve natural language understanding and generation. LLMs have also been used in emotion recognition, dialog act classification, and utterance generation in negotiation tasks \cite{lin2023toward}. 

Several studies have explored generating backchanneling and expressive behavior for social robots. For example, generating robotic backchannelling behaviors have been explored by \cite{mahadevan2024generative}, who used few-shot chain-of-thought prompting to translate human language instructions into parameterized control code.LLMs have also been utilized to generate expressive reactions and mapped them to a robot behavior using a fine-tuned model to be shown on a social robot, leading to more engaging and empathetic interactions \cite{wang2024ain}.

\section{Computational Model of Multimodal Empathic Grounding}
\label{sec:empathic}


We assume a structured, agent-initiated dialog, in which each discourse segment \cite{Grosz1986} is comprised of 1) an agent's question to the user, 2) the user's response to the question, and 3) a grounding response by the agent (\autoref{fig:loop}). Both user and agent contributions to the dialog can be multimodal, comprised of verbal utterances and/or emotional state(s) exhibited during the contribution. Our computational model is illustrated in \autoref{fig:teaser}. The individual components of the model are described below. 

\subsection{Multimodal User Affect Sensing and Recognition}
At each turn of the conversation, the first step of the model is to capture the user's visual and speech input using a camera and microphone (sensing). The user's facial expressions and verbal utterances are derived by processing the visual and speech inputs (recognition). The facial recognition input is then summarized and converted into two most commonly-appeared emotion labels. These processed inputs are then fed into the Interpretation module to generate empathic grounding actions.

\subsection{Multimodal Interpretation} Previous work has demonstrated the effectiveness of LLMs in understanding the emotional tone of text \cite{tak2023gpt, broekens2023fine}. We expand this approach and prompt an LLM as a multimodal emotion interpretation model that captures the underlying emotional state of the user based on their verbal utterances and facial expressions at each turn. Thus, the LLM can be prompted to use the agent's question and the user's multimodal response to predict the user's dominant emotion label (i.e., single-class classification) as interpreted from combined verbal and visual channels in the present turn, along with the valance, arousal, and dominance components of their emotional state \cite{russell1977evidence}.

\subsection{Multimodal Empathic Grounding Generation}
After predicting the user's dominant emotion, the LLM can be prompted to generate an appropriate empathic grounding move. Unlike previous rule-based approaches in affective computing (e.g., \cite{lin2023toward}), which often provide explicit rules for mapping emotional states to behavior generation, we do not define explicit mapping rules and leave decision-making to the LLM to select among multiple available emotional states (e.g., neutral, sad, and happy). Likewise, a set of body movements (e.g., head-nods) can be provided as explicit option sets for the LLM to choose from. Finally, the LLM generates an appropriate verbal utterance.

\begin{figure}[]
    \centering
    \includegraphics[width=0.8\columnwidth]{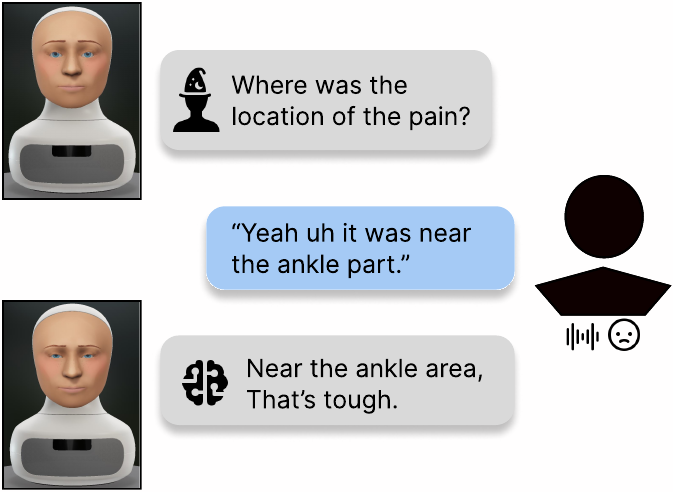} 
    \caption{Example Discourse Segment. 1. Agent question. 2. User response. 3. Agent grounding move.}
    \label{fig:loop}
\end{figure}

\subsection{Implementation Details}
We used the LibreFace Model \cite{chang2023libreface} in the OpenSense application \cite{10.1145/3382507.3418832} to extract user facial expression features. This model provides eight categorical facial expressions corresponding to neutral, surprise, fear, happiness, sadness, disgust, anger, and contempt emotional states. We export labels corresponding to the user's dominant facial expressions at a rate of 15 frames per second and filter out noise using a maximum pool window for every five labels. We then take the two most common labels (excluding neutral) from each user utterance and use them in our multimodal interpreter. 

We used \textit{``gpt-3.5-turbo-0125''} (April 2024 version) as the LLM for the multimodal empathic grounding module. We implemented both the multimodal interpreter and generator in one zero-shot prompt. The input to and output from the LLM were structured in JSON format, which was then parsed by the agent for behavior realization. We instructed the LLM not to give medical advice or inappropriate recommendations, and we did not define a persona for the agent, unlike prior work on affective listeners \cite{mckeown2011semaine}. More comprehensive details of the prompt can be found in \autoref{tab:prompt}, and the complete prompt is included in the supplementary material \footnote{\url{\repourl}}.

As input, the LLM received the agent's question (text only) and the user's multimodal response. Our implementation of the LLM prompt outputs four components for the generated agent behavior (grounding move): 1) emotional state, 2) head movement, 3) verbal utterance, and 4) an explanation for its decisions. Below is an overview of these components:

\begin{itemize}
    \item\textit{Emotional State:} We provided the LLM with a list of possible emotional states for the agent that fit the pain assessment scenario. These include the agent being neutral, sad, happy, concerned, or surprised.
    \item \textit{Head Movement:} The LLM chose from among `no\_movement' and ``head-nod'' actions for agent movements.
    \item \textit{Verbal Utterance:} The LLM was instructed to generate a brief empathic response to the user's action. We prompted LLM to generate a short non-generic utterance to make the user feel heard and reflect on their emotion in an appropriate way. We defined some rules for not generating any medical recommendations and not asking any questions from the user.
    \item \textit{Explanation:} We requested that LLM clarify the reason for choosing its outputs.
\end{itemize}

\begin{table*}[htbp]
    \centering
    \caption{Elements of LLM Prompt Used to Generate Empathic Grounding Moves}
    \label{tab:prompt}
      \resizebox{.85\textwidth}{!}{
    \begin{tabular}{p{0.15\textwidth}|p{0.35\textwidth}|p{0.45\textwidth}}
        
        \textbf{Subsection} & \textbf{Description} & \textbf{Part of the prompt} \\
        \hline
        \hline
        Defining the role &
            Describes the role of the LLM as a Multimodal grounding generator. &
            \textit{"You are performing multimodal grounding moves and back-channeling(BC) for a conversational agent. For input, you will receive..."} \\
        \hline
        Input channel descriptions &
            Details of each value input to or output from the LLM, including input channels, output options, and rules. &
            \textit{"Description of the input channels: \{ "agent\_utterance": "the question that has been asked from the user", ...\}"} \\
        \hline
        Multimodal Affect Recognition &
            Instructs LLM to use both input modalities to predict the user's current affective state and generate the associated emotion label and circumplex scores. &
            \textit{"It is only possible to understand users' meaning by focusing on both their verbal and facial expressions...facial expression can be noisy...predict user's dominant emotion and give VAD scores..."} \\
        \hline
        Multimodal Interpretation &
            Use the user's predicted dominant emotion and generate an appropriate response. &
            \textit{"Based on the user dominant emotion you guessed, you need to generate a proper multimodal and empathic grounding move..."} \\
        \hline
        Human-agent conversation description &
            Describes the structure of the discourse segments used in the model. &
            \textit{"Agent:agent\_utterance\_1, User: user\_utterance\_1, Agent: \textless{}Your generated moves, including verbal and non-verbal parts\textgreater{}, Agent: agent\_utterances\_2, ..."} \\
        \hline
        Rules &
            Rules that LLM needs to use in general include the following 1) options, 2) generation rules, and 3) format rules. &
            \textit{"RULES: 1) For the channels you receive options for {[}affect and movements{]}, only choose your response between the provided options ....2)..."} \\
        \hline
        Input and output format &
            Illustrates example JSON inputs and outputs.  &
            \textit{"input format: \{"agent\_utterance":..., "BC\_verbal\_rules": ..., "BC\_nonverbal\_options": ...\}, output format \{"user\_dominant\_emotion": "Classify the user's emotional state into..."\}"} \\
        
    \end{tabular}
}
\end{table*}

\section{Evaluation Testbed}

We developed a testbed to evaluate the empathic grounding model (\autoref{fig:interaction}). Our task domain involves
interviewing participants about a prior painful experience, inspired by \cite{Murali2023}. 
This domain was chosen because it can involve an extended dialog to probe different aspects of pain (e.g., intensity, location, temporal pattern, etc.) and provides many opportunities for 
empathic grounding. There are validated self-report measures that user responses can be validated against, and painful incidents are relatively common, facilitating study recruitment. 

The overall conversation comprised a greeting, a few turns of social chat, an initial open-ended question about the user's pain experience, a series of follow-up questions probing different aspects of pain (some open-ended and some closed-ended), and a farewell. All dialog is agent-initiated, and grounding moves are generated for every user response, including social chat. 

In our testbed, only grounding responses by the agent were fully automated. The agent's questions and task-based follow-up utterances (e.g., "Tell me more.", "Could you repeat that?") were controlled by a human confederate in a wizard-of-oz setup to ensure human-level understanding of the task-related parts of the conversation. This strategy ensured that errors in understanding user task responses did not derail the conversation and lead to poor perceptions of the agent, even if its grounding moves were flawless. In this setup, the agent grounding moves fit into a very constrained role in the dialog, allowing the effects of different grounding models to be isolated and compared.

\label{sec:conv-strc}

\subsection{Embodied Converational Agent}
We used a Furhat humanoid robot \cite{al2012furhat} as our embodied conversational agent, which is a human-like head capable of expressing verbal utterances, facial expressions, gaze, and head movements. Furhat uses a projector to show an animated face and sensors that can track the user's head movements. This setup is optimal for our empathic grounding investigation, as it focuses the user's attention on the robot's face during the interaction. Furhat is equipped with a Google-based automatic speech recognition (ASR) model to listen to users' speech. Realistic speech is generated using Amazon Polly Neural Voices and includes automatic lip-synchronization. We used a female's face and voice as our robot character. For non-verbal behaviors, we used the robot's default surprise and sad faces and created new implementations of happy and concerned facial expressions and head nods.


\subsection{Wizard-of-Oz Framework}
\label{sec:wiz}
In order to provide accurate dialog management and understanding of the effect of empathic grounding moves in a conversation, we designed a Wizard-of-Oz control setup (\autoref{fig:loop}). The wizard hears and sees the participant's interaction with the robot via closed-circuit video and can initiate the next pain assessment question by pushing a button on a control user interface.  After a user responds to the question, and the robot's grounding move is automatically produced and performed, the wizard can decide to either proceed to the next question or to perform other actions as described in \autoref{tab:wizard}. However, the wizard does not control the robot's grounding responses, emotional states, or head movements, as the testbed automatically generates these responses using the generation module.

\begin{table}[htbp]
\small
\caption{Wizard-of-Oz Control Actions}
\label{tab:example}
\begin{tabularx}{\columnwidth}{@{}lX@{}}
\toprule
\textbf{Button} & \textbf{Action and Descriptions} \\
\midrule
User Repeat Response &
Ask the user to repeat a response if the robot could not capture their utterance. \\
\midrule
Interrupt &
\begin{tabular}[t]{@{}l@{}}
If robot interrupts, apologize to the user and\\ encourage them to continue. 
\end{tabular} \\
\midrule
Irrelevant &
\begin{tabular}[t]{@{}l@{}}
Tells the user that the robot cannot answer \\ their question.
\end{tabular} \\
\midrule
Listen-only &
\begin{tabular}[t]{@{}l@{}}
Lets the robot listen to the user's\\ speech without saying anything.
\end{tabular} \\
\bottomrule
\label{tab:wizard}
\end{tabularx}
\end{table}

\section{Comparative Evaluation Experiment}

We conducted a between-subjects experiment to assess the effect of our agent's grounding moves on users while they expressed their recent episodes of pain, comparing our empathic grounding model to a baseline backchannel-only model.
In this setup, we hypothesized that using empathic grounding would help the user feel more heard and build more trust with the agent, compared to non-empathic grounding.
We designed a between-subjects experiment comparing the following conditions: 
\begin{itemize}
    \item \textbf{BACKCHANNEL (acknowledgment):} In this condition, the agent performs non-empathic grounding moves. The agent a neutral facial expression, a head movement randomly chosen from among ``\textit{head-nod}'' and ``\textit{no\_movement}'', as well as an acknowledgment utterance selected at random from a set of neutral grounding utterances, such as "\textit{Noted.}" or "\textit{OK.}", representing a clinical standard of care in pain assessment interviews based on our analysis of online mock interactions.
    \item \textbf{EMPATHIC GROUNDING:} In this condition, the agent performs empathic grounding moves to react to the user's responses based on the model described in \autoref{sec:empathic}.
\end{itemize}

In both conditions, the wizard controls the task conversation flow and encourages the user to keep sharing their thoughts until they stop speaking or explicitly mention they have nothing to add. 

\subsection{Recruitment}
We recruited participants via an online job posting site and social media within our institution. We had three recruitment criteria: over 18 years old, a native speaker of English, and having experienced an episode of pain in the past two weeks. The study was approved by our Institutional Review Board, and participants were compensated for their time and transportation costs.

\subsection{Measures}
\label{sec:measures}
We captured the participant's speech and visual interaction with the robot using an RGB camera and microphone using the OpenSense \cite{10.1145/3382507.3418832} platform in the Microsoft \textbackslash psi data format \cite{bohus2021platform}. 

In addition to sociodemographics, we collected the following self-report measures from participants following their interaction with the robot:  
the Active-Empathic Listening Scale (AELS) \cite{bodie2011active} and  
the Responsiveness of Other Scale \cite{utami2019collaborative} to assess perceptions of the robot's empathy, 
the MSCEIT-Based Perceived Emotional Intelligence Questionnaire (MSCEIT) \cite{mayer2003measuring}, 
the Bond subscale of the Working Alliance Inventory (WAI) \cite{horvath1989development} to assess trust and confidence in working with the robot,
the Godspeed Agent Attitude Questionnaires (GS) \cite{bartneck2009measurement}, 
and a subset of the General Agent Rating items \cite{Bowman} (\autoref{tab:satisfaction}).

Additionally, in order to assess the validity of pain reports to the robot, participants provided self-reports on their pain experiences using validated measures, including PROMIS \cite{bevans2014patient} and the Visual Analog Scale \cite{langley1985visual}.

Finally, semi-structured interviews were conducted with participants at the end of their session to obtain an overall understanding of their experience. Interviews were recorded, transcribed, and analyzed using thematic analysis techniques \cite{braun2006using}.

\begin{table*}
\resizebox{0.9\textwidth}{!}{
\begin{tabular}{|l|l|l|c|c|c|}
\hline
\textbf{Question}                      & \textbf{Anchor 1} & \textbf{Anchor 7}  & \textbf{BACKCHANNEL}
                       & \textbf{EMPATHIC} & \textbf{p} \\ \hline
How satisfied are you with the conversational agent?    & Not at all          & Very satisfied & 4.00 (0.93)  & 5.11 (1.54)  & 0.074           \\ \hline
How much would you like to continue working with the conversational agent?  & Not at all  & Very much    & 4.00 (0.76) & 5.56 (1.24) & 0.015$^{*}$      \\ \hline
How much do you like the conversational agent?      & Not at all          & Very much   & 3.63 (1.06) & 5.78 (1.39) & 0.006$^{**}$         \\ \hline
How would you characterize your relationship with the agent?                 &Stranger & Close friend   & 2.88 (1.36) & 4.22 (1.92) & 0.139          \\ \hline
How much do you feel the conversational agent cares about you? & Not at all   & Very    & 3.50 (2.27) & 5.00 (1.50) & 0.200             \\ \hline
How much do you feel that you and the conversational agent understand each other? & Not at all & Very much    & 3.13 (1.36) & 4.78 (1.39) & 0.036$^{*}$         \\ \hline
\end{tabular}}
\caption{Self-Report Attitudes towards Robot by Study Condition. Wilcoxon significance tests on single items.}
\label{tab:satisfaction}
\end{table*}

\subsection{Study Protocol} 
Following informed consent, we randomly assigned participants to one of our two groups (BACKCHANNEL and EMPATHIC GROUNDING). 
We then introduced the robot as an automated agent that can see and hear people and ask them about a recent painful experience. Participants were informed that they should respond to the robot's questions with natural speech. A researcher (wizard) controlled the robot from a different room, using the method described in \autoref{sec:wiz}. 

Following the pain interview with the robot, participants filled out post-study questionnaires and participated in the semi-structured interview \autoref{sec:measures}.
Finally, participants received a debriefing about the system and the WoZ deception for controlling the robot. 

\subsection{Results}

\subsubsection{Participants} 
A total of 18 participants completed the experiment, including 12 males and 6 females.  Participants were aged between 21 and 64 (Med =30, IQR=23.5). In terms of race, 50\% of participants identified as White, 33.3\% as Asian, and 5.5\% as African American, while others were multi-racial or preferred not to disclose.  Most had completed higher education (high school=2, bachelor's degree=7, graduate degree=9) and had experience with computers (regularly using computer=5, expert using computer=13).  Lastly, participants reported the intensity of their reported pain experience an average of $\mu = 5.32, \sigma=1.61$ on the Visual Analog Scale (ranging from 0 for no pain to 10 as ``worst pain''). We excluded one participant's data due to a technical error that led to incomplete interaction.

\subsection{Quantitative Analysis} 

\begin{figure}[htbp]
    \centering
    \includegraphics[width=0.8\columnwidth]{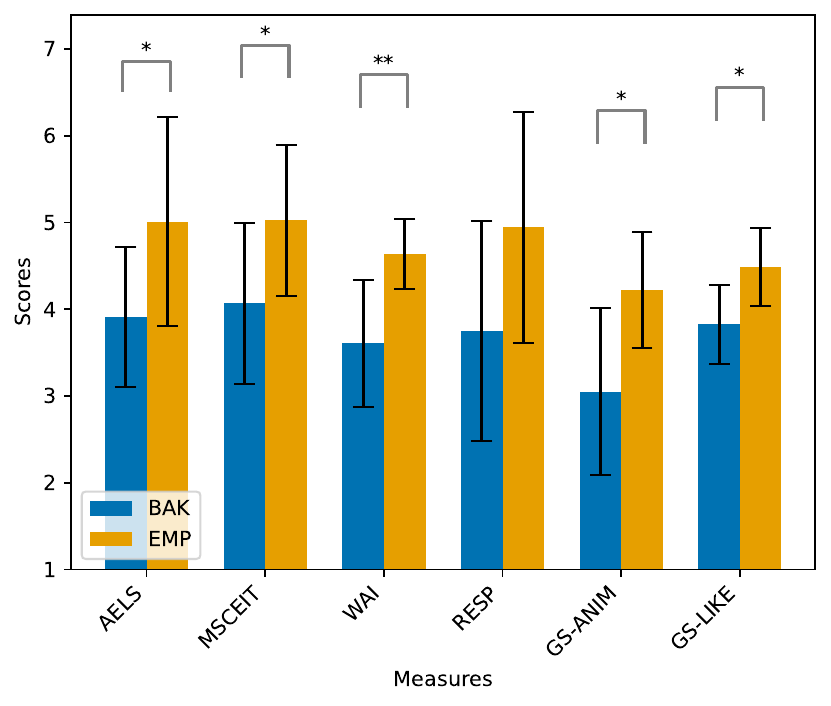}
    \caption{Ratings of BACKCHANNEL and EMPATHIC GROUNDING Robots. See \autoref{sec:measures} for other measure descriptions. (*) shows $p<0.05$ and (**) shows $p<0.01$}
    \label{fig:measures}
\end{figure}

Although we had a small sample size, Shapiro-Wilk normality tests showed the normality of the distributions of our measurements WAI, MSCEIT, AELS, and Responsiveness. As demonstrated in \autoref{fig:measures}, we performed an independent sample t-test for composite measures between two conditions.   Additionally, we compared the single-item measures of satisfaction using Wilcoxon (\autoref{tab:satisfaction}.)

Compared to those in the BACKCHANNEL condition, participants in the EMPATHIC GROUNDING condition rated the robot significantly higher on empathy, based on AELS, $t(15)=2.173, p<.05$, significantly higher on emotional intelligence, based on MSCEIT, $t(15)=2,186, p<.05$, and significantly higher on working alliance, based on WAI-Bond, $t(13)=3.438, p<0.01$. Differences in Responsiveness (RESP) were in the hypothesized direction but were only trending in significance, $t(15)=1.886, p=.079$. On the Godspeed measures, the EMPATHIC GROUNDING robot was found to be significantly more animated (GS-ANIM), $t(16)=2.53, p=.022$, and likable (GS-LIKE), $t(16)=2.18, p=.045$, compared to the BACKCHANNEL robot. These results are illustrated in \autoref{fig:measures}.

Based on Wilcoxon tests on the General Agent Rating items (\autoref{tab:satisfaction}), participants in the EMPATHIC GROUNDING condition rated the robot significantly higher on the degree they felt the robot understood them, satisfaction with the robot, desire to continue working with the robot, and liking of the robot.

We checked the validity of participant pain reports to the robot against the PROMIS and Visual Analog Scale measures. Those in the EMPATHIC GROUNDING group reported pain intensity that was significantly correlated with a numeric scale of pain intensity, Spearman $rho=0.932, p<.001$, but whose correlation with the Visual Analog Scale of pain intensity was only trending, Spearman $rho=0.635, p=.066$. Those in the BACKCHANNEL group reported pain intensity that was significantly correlated with both standard measures, Spearman $rho=0.897, p=.003$ for the numeric scale, and Spearman $rho=0.830, p=.011$ for the Visual Analog Scale. 

For reports of the degree to which pain was interfering with their life, those in the EMPATHIC GROUNDING group reported interference that was significantly correlated with the corresponding PROMIS measure, Spearman $rho=0.805, p=.016$, while for those in the BACKCHANNEL group, this correlation was not significant, Spearman $rho=0.506, p=.201$.

\subsection{Qualitative Analysis}
From semi-structured interviews, we analyzed a total of 171 minutes of transcripts using thematic analysis to understand user perceptions of the interaction with the robot. The most prominent themes that emerged from the data are described below.

\subsubsection{Perceptions of Empathic Grounding} Participants in the EMPATHIC GROUNDING group (8 out of 9 participants) commented on features explicitly related to the empathic grounding moves in their feedback. Some liked the expressions and noticed instances of facial expressions and emotions: "\textit{When it was saying something very sad, it [looked] down...It understands and gives a sad expression.}" [P12]. Another participant highlighted: \textit{"it was sensing my emotion in whatever I was [saying], it was [interpreting] the context and if I speak positively, negatively, sadly happily it was getting that"}. However, some found these responses partially scripted: "\textit{There were a couple of times I thought that it did [listen to me]...But for the majority of the time it [was] more pre-programmed, scripted response.}" [P4].

\subsubsection{Feeling heard by the robot} 
Only 3 out of 8 participants in our BACKCHANNEL group mentioned being heard by the agent, while 9 out of 9 participants in EMPATHIC GROUNDING said the robot listened to them during the conversation. Although in the BACKCHANNEL group, the robot acknowledged what the participant had mentioned, some participants felt that the interaction was programmed: "\textit{It listened. I mean it recorded me. I didn't really get the feeling it was listening as much as recording}" [P15, BACKCHANNEL]. This was in contrast to most participants in EMPATHIC GROUNDING, who felt that the robot genuinely heard them. Additionally, multiple participants in this group emphasized that the interaction felt natural, possibly contributing to the feeling of being heard. 

\subsubsection{Being interrupted by the Robot}
The majority of our participants (6 out of 8) in the BACKCHANNEL group and 4 out of 9 participants in the EMPATHIC GROUNDING group mentioned that the robot interrupted them in the middle of their conversation. Although we used the same robot setup in both groups, the rate of reporting the interruption during the open-ended interviewing was different.

\subsection{Case Analysis: LLM as Multimodal Reasoner}
Throughout the experiment, we observed several interactions in which information from the user's affective state, as recognized from their facial display, was used to provide more empathically-accurate grounding than would have been otherwise possible for those in the EMPATHIC GROUNDING condition.
For example, P16's response to the question "\textit{How's the weather today?}" was "\textit{cloudy.}" with a smiling facial display, and the robot used the empathic grounding move: "\textit{I'm glad you find joy in it}".
The LLM's explanation for this action was "\textit{The user's verbal response 'cloudy' combined with facial expressions of 'Happiness' indicate a positive and cheerful mood}".
Based on a post-experiment analysis, we found that had the user affect information not been provided to the LLM, it would have generated the more neutral sentence "\textit{It looks like a cloudy day}."

Another example was during P14's social chat. When asked about the weather, they responded, "\textit{it's raining outside and sometimes it is showing to be snowy and I like it personally but it is a little bit uh not that comfortable but it's all right}" while smiling. In response, the agent's reaction was a head-nod, happy face, and a "\textit{It sounds like a lovely mix!}" utterance. However, if user affect information had been removed, the response would have been an agent with a neutral face saying, "\textit{Sounds like a mixed weather day}".

As part of the conversation with P18's interaction, when the agent asked if they had experienced a painful experience recently, they said, "\textit{that is correct}" while smiling. The agent's grounding move was "\textit{I appreciate your honesty and strength}" with a neutral face and a head nod. The LLM's rationale for this was "The user confirmed a painful experience but expressed happiness, indicating a resilient attitude, so a supportive and appreciative response is suitable". However, if LLM hadn't used the facial expression, the generated sentence would have been "\textit{I understand your pain. Take care}" with a concerned facial expression and no head movement.

Participants in the EMPATHIC GROUNDING condition indicated that they and the robot understood each other to a significantly greater degree compared to those in the BACKCHANNEL condition, supporting \textbf{H1}. Participants in the EMPATHIC GROUNDING condition also rated the robot significantly higher on the Active-Empathic Listening Scale (AELS), supporting \textbf{H2}. Participants in the EMPATHIC GROUNDING condition also rated the robot significantly higher on the Bond dimension of the Working Alliance Inventory, indicating a higher degree of trust, supporting \textbf{H3}. Finally, we identified several instances in which our empathic grounding model incorporated information about the user's affective state to produce responses that were more appropriate than they would have been otherwise, providing partial support for \textbf{H4}. That said, a more systematic approach is required to fully investigate the impact of multimodal input channels on the quality of empathic grounding moves.

Participants in the EMPATHIC GROUNDING condition also rated the robot significantly higher on Perceived Emotional Intelligence. This scale has items spanning the identification of emotions, appropriate responses to emotions, and the appearance of being emotionally self-aware. 

Our analysis suggests that empathic grounding can alter the user's perception of the interaction and make them feel heard more often. Not every participant in our EMPATHIC GROUNDING study felt their emotions were being used in the same way, but some noticed that the agent was able to convey emotions in accordance with what they had expressed. 

Interestingly, even in conversational exchanges that were not intended to require empathic grounding--such as social chat about the weather--the empathic grounding model still provided more appropriate and affectively-tailored responses to user utterances, compared to a model that only responded with neutral backchannel responses.

\section{Conclusion}
We introduced the concept of "empathic grounding" in conversational agents, and described how it fits into existing theories of macro- and micro-empathy, as well as grounding in face-to-face communication. We described a model for automatically generating multimodal empathic grounding responses for an embodied conversational agent using LLMs, and a testbed for evaluating such models in the task domain of interviewing people about past pain experiences. Finally, we presented results from an experiment, demonstrating that our empathic grounding model leads to significantly higher ratings of empathic listening, emotional intelligence, and working alliance for an agent compared to an agent that only uses neutral backchannels for grounding.

\subsection{Limitations}
There are many limitations to our study, beyond the small convenience sample used. We evaluated our empathic grounding model in a highly constrained scenario, and our results may not generalize to other kinds of conversations or even other dialog structures.

The two conditions we compared in our study may not have been ideal for isolating a single factor for testing.
However, we feel they represent a pragmatic trial in which the BACKCHANNEL condition is a good representation of
'standard of care' in clinical pain assessment, and the EMPATHIC-GROUNDING condition reflects the behavior of
an affectively-attuned, empathic caregiver.


\subsection{Future Work}
There are many interesting directions for future work. Grounding includes much more than providing positive evidence of understanding, and the use of LLMs to generate other kinds of grounding moves--including negative evidence and repairs--remains an important avenue of investigation. We only used participant facial display for affect recognition, and there are many other modalities and approaches that could provide more accurate and nuanced assessments that could be included. 

We only used one turn of conversation to generate empathic grounding moves, but future work can assess the role of including more discourse context in move generation. 

Due to a fixed audio threshold timeout to determine end of user utterance in Furhat, the robot frequently interrupted users who paused during long contributions. Even though the robot interrupted with a grounding move, it often disrupted and annoyed users. Future work should explore the use of better turn-taking mechanisms, including the incorporation of multimodal cues such as user gaze, as well as a more nuanced generation of grounding moves that are sensitive to whether the user is trying to hold the floor or not (e.g., using only backchannels during pauses in long user contributions). 

Finally, the nonverbal cues used to express empathy can be culturally-dependent \cite{Lorie2017}, raising the question of how grounding differs across cultures, and this represents another important area of inquiry and experimentation.

\bibliographystyle{ACM-Reference-Format}
\bibliography{sample-base}


\begin{thebibliography}{62}


\ifx \showCODEN    \undefined \def \showCODEN     #1{\unskip}     \fi
\ifx \showDOI      \undefined \def \showDOI       #1{#1}\fi
\ifx \showISBNx    \undefined \def \showISBNx     #1{\unskip}     \fi
\ifx \showISBNxiii \undefined \def \showISBNxiii  #1{\unskip}     \fi
\ifx \showISSN     \undefined \def \showISSN      #1{\unskip}     \fi
\ifx \showLCCN     \undefined \def \showLCCN      #1{\unskip}     \fi
\ifx \shownote     \undefined \def \shownote      #1{#1}          \fi
\ifx \showarticletitle \undefined \def \showarticletitle #1{#1}   \fi
\ifx \showURL      \undefined \def \showURL       {\relax}        \fi
\providecommand\bibfield[2]{#2}
\providecommand\bibinfo[2]{#2}
\providecommand\natexlab[1]{#1}
\providecommand\showeprint[2][]{arXiv:#2}

\bibitem[Achiam et~al\mbox{.}(2023)]%
        {achiam2023gpt}
\bibfield{author}{\bibinfo{person}{Josh Achiam}, \bibinfo{person}{Steven Adler}, \bibinfo{person}{Sandhini Agarwal}, \bibinfo{person}{Lama Ahmad}, \bibinfo{person}{Ilge Akkaya}, \bibinfo{person}{Florencia~Leoni Aleman}, \bibinfo{person}{Diogo Almeida}, \bibinfo{person}{Janko Altenschmidt}, \bibinfo{person}{Sam Altman}, \bibinfo{person}{Shyamal Anadkat}, {et~al\mbox{.}}} \bibinfo{year}{2023}\natexlab{}.
\newblock \showarticletitle{Gpt-4 technical report}.
\newblock \bibinfo{journal}{\emph{arXiv preprint arXiv:2303.08774}} (\bibinfo{year}{2023}).
\newblock


\bibitem[Al~Moubayed et~al\mbox{.}(2012)]%
        {al2012furhat}
\bibfield{author}{\bibinfo{person}{Samer Al~Moubayed}, \bibinfo{person}{Jonas Beskow}, \bibinfo{person}{Gabriel Skantze}, {and} \bibinfo{person}{Bj{\"o}rn Granstr{\"o}m}.} \bibinfo{year}{2012}\natexlab{}.
\newblock \showarticletitle{Furhat: a back-projected human-like robot head for multiparty human-machine interaction}. In \bibinfo{booktitle}{\emph{Cognitive Behavioural Systems: COST 2102 International Training School, Dresden, Germany, February 21-26, 2011, Revised Selected Papers}}. Springer, \bibinfo{pages}{114--130}.
\newblock


\bibitem[Allwood et~al\mbox{.}(1992)]%
        {Allwood1992}
\bibfield{author}{\bibinfo{person}{J Allwood}, \bibinfo{person}{J Nivre}, {and} \bibinfo{person}{E Ahlsen}.} \bibinfo{year}{1992}\natexlab{}.
\newblock \showarticletitle{On the semantics and pragmatics of linguistic feedback}.
\newblock \bibinfo{journal}{\emph{Journal of Semantics}}  \bibinfo{volume}{9} (\bibinfo{year}{1992}).
\newblock


\bibitem[Ayers et~al\mbox{.}(2023)]%
        {Ayers2023}
\bibfield{author}{\bibinfo{person}{J.~W. Ayers}, \bibinfo{person}{A. Poliak}, \bibinfo{person}{M. Dredze}, \bibinfo{person}{E.~C. Leas}, \bibinfo{person}{Z. Zhu}, \bibinfo{person}{J.~B. Kelley}, \bibinfo{person}{D.~J. Faix}, \bibinfo{person}{A.~M. Goodman}, \bibinfo{person}{C.~A. Longhurst}, \bibinfo{person}{M. Hogarth}, {and} \bibinfo{person}{D.~M. Smith}.} \bibinfo{year}{2023}\natexlab{}.
\newblock \showarticletitle{Comparing Physician and Artificial Intelligence Chatbot Responses to Patient Questions Posted to a Public Social Media Forum}.
\newblock \bibinfo{journal}{\emph{JAMA Intern Med}} \bibinfo{volume}{183}, \bibinfo{number}{6} (\bibinfo{year}{2023}), \bibinfo{pages}{589--596}.
\newblock


\bibitem[Bartneck et~al\mbox{.}(2009)]%
        {bartneck2009measurement}
\bibfield{author}{\bibinfo{person}{Christoph Bartneck}, \bibinfo{person}{Dana Kuli{\'c}}, \bibinfo{person}{Elizabeth Croft}, {and} \bibinfo{person}{Susana Zoghbi}.} \bibinfo{year}{2009}\natexlab{}.
\newblock \showarticletitle{Measurement instruments for the anthropomorphism, animacy, likeability, perceived intelligence, and perceived safety of robots}.
\newblock \bibinfo{journal}{\emph{International journal of social robotics}} \bibinfo{volume}{1}, \bibinfo{number}{1} (\bibinfo{year}{2009}), \bibinfo{pages}{71--81}.
\newblock


\bibitem[Bavelas et~al\mbox{.}(2000)]%
        {bavelas2000listeners}
\bibfield{author}{\bibinfo{person}{Janet~B Bavelas}, \bibinfo{person}{Linda Coates}, {and} \bibinfo{person}{Trudy Johnson}.} \bibinfo{year}{2000}\natexlab{}.
\newblock \showarticletitle{Listeners as co-narrators.}
\newblock \bibinfo{journal}{\emph{Journal of personality and social psychology}} \bibinfo{volume}{79}, \bibinfo{number}{6} (\bibinfo{year}{2000}), \bibinfo{pages}{941}.
\newblock


\bibitem[Bertrand et~al\mbox{.}(2007)]%
        {Bertrand2007}
\bibfield{author}{\bibinfo{person}{R Bertrand}, \bibinfo{person}{G Ferré}, \bibinfo{person}{P Blache}, \bibinfo{person}{R Espesser}, {and} \bibinfo{person}{S Rauzy}.} \bibinfo{year}{2007}\natexlab{}.
\newblock \bibinfo{booktitle}{\emph{Backchannels revisited from a multimodal perspective}}.
\newblock \bibinfo{pages}{1--5}.
\newblock


\bibitem[Bevans et~al\mbox{.}(2014)]%
        {bevans2014patient}
\bibfield{author}{\bibinfo{person}{Margaret Bevans}, \bibinfo{person}{Alyson Ross}, {and} \bibinfo{person}{David Cella}.} \bibinfo{year}{2014}\natexlab{}.
\newblock \showarticletitle{Patient-Reported Outcomes Measurement Information System (PROMIS): efficient, standardized tools to measure self-reported health and quality of life}.
\newblock \bibinfo{journal}{\emph{Nursing outlook}} \bibinfo{volume}{62}, \bibinfo{number}{5} (\bibinfo{year}{2014}), \bibinfo{pages}{339--345}.
\newblock


\bibitem[Bilalpur et~al\mbox{.}(2024)]%
        {Bilalpur2024}
\bibfield{author}{\bibinfo{person}{M Bilalpur}, \bibinfo{person}{M Inan}, \bibinfo{person}{D Zeinali}, \bibinfo{person}{J Cohn}, {and} \bibinfo{person}{M Alikhani}.} \bibinfo{year}{2024}\natexlab{}.
\newblock \bibinfo{title}{Learning to Generate Context-Sensitive Backchannel Smiles for Embodied AI Agents with Applications in Mental Health Dialogues}.
\newblock
\newblock
\urldef\tempurl%
\url{https://arxiv.org/html/2402.08837v1}
\showURL{%
\tempurl}


\bibitem[Bodie(2011)]%
        {bodie2011active}
\bibfield{author}{\bibinfo{person}{Graham~D Bodie}.} \bibinfo{year}{2011}\natexlab{}.
\newblock \showarticletitle{The Active-Empathic Listening Scale (AELS): Conceptualization and evidence of validity within the interpersonal domain}.
\newblock \bibinfo{journal}{\emph{Communication Quarterly}} \bibinfo{volume}{59}, \bibinfo{number}{3} (\bibinfo{year}{2011}), \bibinfo{pages}{277--295}.
\newblock


\bibitem[Bohus et~al\mbox{.}(2021)]%
        {bohus2021platform}
\bibfield{author}{\bibinfo{person}{Dan Bohus}, \bibinfo{person}{Sean Andrist}, \bibinfo{person}{Ashley Feniello}, \bibinfo{person}{Nick Saw}, \bibinfo{person}{Mihai Jalobeanu}, \bibinfo{person}{Patrick Sweeney}, \bibinfo{person}{Anne~Loomis Thompson}, {and} \bibinfo{person}{Eric Horvitz}.} \bibinfo{year}{2021}\natexlab{}.
\newblock \bibinfo{title}{Platform for Situated Intelligence}.
\newblock
\newblock
\showeprint[arxiv]{2103.15975}~[cs.AI]


\bibitem[Bowman et~al\mbox{.}(2024)]%
        {Bowman}
\bibfield{author}{\bibinfo{person}{R Bowman}, \bibinfo{person}{O Cooney}, \bibinfo{person}{J Newbold}, \bibinfo{person}{A Thieme}, \bibinfo{person}{L Clark}, \bibinfo{person}{G Doherty}, {and} \bibinfo{person}{Cowan B}.} \bibinfo{year}{2024}\natexlab{}.
\newblock \showarticletitle{Exploring how politeness impacts the user experience of chatbots for mental health support}.
\newblock \bibinfo{journal}{\emph{International Journal of Human-Computer Studies}}  \bibinfo{volume}{184} (\bibinfo{year}{2024}).
\newblock


\bibitem[Braun and Clarke(2006)]%
        {braun2006using}
\bibfield{author}{\bibinfo{person}{Virginia Braun} {and} \bibinfo{person}{Victoria Clarke}.} \bibinfo{year}{2006}\natexlab{}.
\newblock \showarticletitle{Using thematic analysis in psychology}.
\newblock \bibinfo{journal}{\emph{Qualitative research in psychology}} \bibinfo{volume}{3}, \bibinfo{number}{2} (\bibinfo{year}{2006}), \bibinfo{pages}{77--101}.
\newblock


\bibitem[Broekens et~al\mbox{.}(2023)]%
        {broekens2023fine}
\bibfield{author}{\bibinfo{person}{Joost Broekens}, \bibinfo{person}{Bernhard Hilpert}, \bibinfo{person}{Suzan Verberne}, \bibinfo{person}{Kim Baraka}, \bibinfo{person}{Patrick Gebhard}, {and} \bibinfo{person}{Aske Plaat}.} \bibinfo{year}{2023}\natexlab{}.
\newblock \showarticletitle{Fine-grained Affective Processing Capabilities Emerging from Large Language Models}. In \bibinfo{booktitle}{\emph{2023 11th International Conference on Affective Computing and Intelligent Interaction (ACII)}}. IEEE, \bibinfo{pages}{1--8}.
\newblock


\bibitem[Brown et~al\mbox{.}(2020)]%
        {brown2020language}
\bibfield{author}{\bibinfo{person}{Tom Brown}, \bibinfo{person}{Benjamin Mann}, \bibinfo{person}{Nick Ryder}, \bibinfo{person}{Melanie Subbiah}, \bibinfo{person}{Jared~D Kaplan}, \bibinfo{person}{Prafulla Dhariwal}, \bibinfo{person}{Arvind Neelakantan}, \bibinfo{person}{Pranav Shyam}, \bibinfo{person}{Girish Sastry}, \bibinfo{person}{Amanda Askell}, {et~al\mbox{.}}} \bibinfo{year}{2020}\natexlab{}.
\newblock \showarticletitle{Language models are few-shot learners}.
\newblock \bibinfo{journal}{\emph{Advances in neural information processing systems}}  \bibinfo{volume}{33} (\bibinfo{year}{2020}), \bibinfo{pages}{1877--1901}.
\newblock


\bibitem[Buschmeier and Kopp(2014)]%
        {buschmeier2014elicit}
\bibfield{author}{\bibinfo{person}{Hendrik Buschmeier} {and} \bibinfo{person}{Stefan Kopp}.} \bibinfo{year}{2014}\natexlab{}.
\newblock \showarticletitle{When to elicit feedback in dialogue: Towards a model based on the information needs of speakers}. In \bibinfo{booktitle}{\emph{Intelligent Virtual Agents: 14th International Conference, IVA 2014, Boston, MA, USA, August 27-29, 2014. Proceedings 14}}. Springer, \bibinfo{pages}{71--80}.
\newblock


\bibitem[Carr et~al\mbox{.}(2003)]%
        {Carr2003}
\bibfield{author}{\bibinfo{person}{L. Carr}, \bibinfo{person}{M. Iacoboni}, \bibinfo{person}{M.~C. Dubeau}, \bibinfo{person}{J.~C. Mazziotta}, {and} \bibinfo{person}{G.~L. Lenzi}.} \bibinfo{year}{2003}\natexlab{}.
\newblock \showarticletitle{Neural mechanisms of empathy in humans: a relay from neural systems for imitation to limbic areas}.
\newblock \bibinfo{journal}{\emph{Proc Natl Acad Sci U S A}} \bibinfo{volume}{100}, \bibinfo{number}{9} (\bibinfo{year}{2003}), \bibinfo{pages}{5497--502}.
\newblock


\bibitem[Cassell et~al\mbox{.}(2001)]%
        {cassell2001beat}
\bibfield{author}{\bibinfo{person}{Justine Cassell}, \bibinfo{person}{Hannes~H{\"o}gni Vilhj{\'a}lmsson}, {and} \bibinfo{person}{Timothy Bickmore}.} \bibinfo{year}{2001}\natexlab{}.
\newblock \showarticletitle{Beat: the behavior expression animation toolkit}. In \bibinfo{booktitle}{\emph{Proceedings of the 28th annual conference on Computer graphics and interactive techniques}}. \bibinfo{pages}{477--486}.
\newblock


\bibitem[Chang et~al\mbox{.}(2024)]%
        {chang2023libreface}
\bibfield{author}{\bibinfo{person}{Di Chang}, \bibinfo{person}{Yufeng Yin}, \bibinfo{person}{Zongjian Li}, \bibinfo{person}{Minh Tran}, {and} \bibinfo{person}{Mohammad Soleymani}.} \bibinfo{year}{2024}\natexlab{}.
\newblock \showarticletitle{LibreFace: An Open-Source Toolkit for Deep Facial Expression Analysis}. In \bibinfo{booktitle}{\emph{Proceedings of the IEEE/CVF Winter Conference on Applications of Computer Vision (WACV)}}.
\newblock
\newblock
\shownote{To appear}.


\bibitem[Clark and Brennan(1991)]%
        {Clark1991}
\bibfield{author}{\bibinfo{person}{Herbert~H. Clark} {and} \bibinfo{person}{Susan~E. Brennan}.} \bibinfo{year}{1991}\natexlab{}.
\newblock \bibinfo{booktitle}{\emph{Grounding in Communication}}.
\newblock \bibinfo{publisher}{American Psychological Association}, \bibinfo{address}{Washington}, \bibinfo{pages}{127--149}.
\newblock
\newblock
\shownote{read}.


\bibitem[Cuadra et~al\mbox{.}(2024)]%
        {cuadra2024illusion}
\bibfield{author}{\bibinfo{person}{Andrea Cuadra}, \bibinfo{person}{Maria Wang}, \bibinfo{person}{Lynn~Andrea Stein}, \bibinfo{person}{Malte~F Jung}, \bibinfo{person}{Nicola Dell}, \bibinfo{person}{Deborah Estrin}, {and} \bibinfo{person}{James~A Landay}.} \bibinfo{year}{2024}\natexlab{}.
\newblock \showarticletitle{The Illusion of Empathy? Notes on Displays of Emotion in Human-Computer Interaction}. In \bibinfo{booktitle}{\emph{ACM Conference on Human Factors in Computing Systems (CHI)}}.
\newblock


\bibitem[DeVault et~al\mbox{.}(2014)]%
        {devault2014simsensei}
\bibfield{author}{\bibinfo{person}{David DeVault}, \bibinfo{person}{Ron Artstein}, \bibinfo{person}{Grace Benn}, \bibinfo{person}{Teresa Dey}, \bibinfo{person}{Ed Fast}, \bibinfo{person}{Alesia Gainer}, \bibinfo{person}{Kallirroi Georgila}, \bibinfo{person}{Jon Gratch}, \bibinfo{person}{Arno Hartholt}, \bibinfo{person}{Margaux Lhommet}, {et~al\mbox{.}}} \bibinfo{year}{2014}\natexlab{}.
\newblock \showarticletitle{SimSensei Kiosk: A virtual human interviewer for healthcare decision support}. In \bibinfo{booktitle}{\emph{Proceedings of the 2014 international conference on Autonomous agents and multi-agent systems}}. \bibinfo{pages}{1061--1068}.
\newblock


\bibitem[Gratch and Lucas(2021)]%
        {10.1145/3477322.3477335}
\bibfield{author}{\bibinfo{person}{Jonathan Gratch} {and} \bibinfo{person}{Gale Lucas}.} \bibinfo{year}{2021}\natexlab{}.
\newblock \bibinfo{booktitle}{\emph{Rapport Between Humans and Socially Interactive Agents} (\bibinfo{edition}{1} ed.)}.
\newblock \bibinfo{publisher}{Association for Computing Machinery}, \bibinfo{address}{New York, NY, USA}, \bibinfo{pages}{433–462}.
\newblock
\showISBNx{9781450387200}
\urldef\tempurl%
\url{https://doi.org/10.1145/3477322.3477335}
\showURL{%
\tempurl}


\bibitem[Gratch et~al\mbox{.}(2006)]%
        {gratch2006virtual}
\bibfield{author}{\bibinfo{person}{Jonathan Gratch}, \bibinfo{person}{Anna Okhmatovskaia}, \bibinfo{person}{Francois Lamothe}, \bibinfo{person}{Stacy Marsella}, \bibinfo{person}{Mathieu Morales}, \bibinfo{person}{Rick~J van~der Werf}, {and} \bibinfo{person}{Louis-Philippe Morency}.} \bibinfo{year}{2006}\natexlab{}.
\newblock \showarticletitle{Virtual rapport}. In \bibinfo{booktitle}{\emph{Intelligent Virtual Agents: 6th International Conference, IVA 2006, Marina Del Rey, CA, USA, August 21-23, 2006. Proceedings 6}}. Springer, \bibinfo{pages}{14--27}.
\newblock


\bibitem[Gratch et~al\mbox{.}(2007)]%
        {gratch2007creating}
\bibfield{author}{\bibinfo{person}{Jonathan Gratch}, \bibinfo{person}{Ning Wang}, \bibinfo{person}{Jillian Gerten}, \bibinfo{person}{Edward Fast}, {and} \bibinfo{person}{Robin Duffy}.} \bibinfo{year}{2007}\natexlab{}.
\newblock \showarticletitle{Creating rapport with virtual agents}. In \bibinfo{booktitle}{\emph{Intelligent Virtual Agents: 7th International Conference, IVA 2007 Paris, France, September 17-19, 2007 Proceedings 7}}. Springer, \bibinfo{pages}{125--138}.
\newblock


\bibitem[Grice(1975)]%
        {Grice}
\bibfield{author}{\bibinfo{person}{H Grice}.} \bibinfo{year}{1975}\natexlab{}.
\newblock \bibinfo{booktitle}{\emph{Logic and conversation}}.
\newblock \bibinfo{publisher}{Academic Press}, \bibinfo{address}{New York}.
\newblock


\bibitem[Grosz and Sidner(1986)]%
        {Grosz1986}
\bibfield{author}{\bibinfo{person}{Barbara Grosz} {and} \bibinfo{person}{C. Sidner}.} \bibinfo{year}{1986}\natexlab{}.
\newblock \showarticletitle{Attention, Intentions, and the Structure of Discourse}.
\newblock \bibinfo{journal}{\emph{Computational Linguistics}} \bibinfo{volume}{12}, \bibinfo{number}{3} (\bibinfo{year}{1986}), \bibinfo{pages}{175--204}.
\newblock


\bibitem[Heylen(2007)]%
        {heylen2007multimodal}
\bibfield{author}{\bibinfo{person}{Dirk Heylen}.} \bibinfo{year}{2007}\natexlab{}.
\newblock \showarticletitle{Multimodal backchannel generation for conversational agents}. In \bibinfo{booktitle}{\emph{Workshop on Multimodal Output Generation, MOG 2007}}. Centre for Telematics and Information Technology (CTIT), \bibinfo{pages}{81--92}.
\newblock


\bibitem[Horvath and Greenberg(1989)]%
        {horvath1989development}
\bibfield{author}{\bibinfo{person}{Adam~O Horvath} {and} \bibinfo{person}{Leslie~S Greenberg}.} \bibinfo{year}{1989}\natexlab{}.
\newblock \showarticletitle{Development and validation of the Working Alliance Inventory.}
\newblock \bibinfo{journal}{\emph{Journal of counseling psychology}} \bibinfo{volume}{36}, \bibinfo{number}{2} (\bibinfo{year}{1989}), \bibinfo{pages}{223}.
\newblock


\bibitem[Huang et~al\mbox{.}(2011)]%
        {huang2011virtual}
\bibfield{author}{\bibinfo{person}{Lixing Huang}, \bibinfo{person}{Louis-Philippe Morency}, {and} \bibinfo{person}{Jonathan Gratch}.} \bibinfo{year}{2011}\natexlab{}.
\newblock \showarticletitle{Virtual Rapport 2.0}. In \bibinfo{booktitle}{\emph{Intelligent Virtual Agents: 10th International Conference, IVA 2011, Reykjavik, Iceland, September 15-17, 2011. Proceedings 11}}. Springer, \bibinfo{pages}{68--79}.
\newblock


\bibitem[Jung(2017)]%
        {Jung}
\bibfield{author}{\bibinfo{person}{M Jung}.} \bibinfo{year}{2017}\natexlab{}.
\newblock \showarticletitle{Affective Grounding in Human-Robot Interaction}. In \bibinfo{booktitle}{\emph{HRI}}.
\newblock


\bibitem[Keen(2010)]%
        {Keen2010}
\bibfield{author}{\bibinfo{person}{S Keen}.} \bibinfo{year}{2010}\natexlab{}.
\newblock \bibinfo{booktitle}{\emph{Narrative empathy}}.
\newblock \bibinfo{publisher}{University of Texas Press}, \bibinfo{pages}{61--94}.
\newblock


\bibitem[Kopp and Hassan(2022)]%
        {10.1145/3563659.3563664}
\bibfield{author}{\bibinfo{person}{Stefan Kopp} {and} \bibinfo{person}{Teena Hassan}.} \bibinfo{year}{2022}\natexlab{}.
\newblock \bibinfo{booktitle}{\emph{The Fabric of Socially Interactive Agents: Multimodal Interaction Architectures} (\bibinfo{edition}{1} ed.)}.
\newblock \bibinfo{publisher}{Association for Computing Machinery}, \bibinfo{address}{New York, NY, USA}, \bibinfo{pages}{77–112}.
\newblock
\showISBNx{9781450398961}
\urldef\tempurl%
\url{https://doi.org/10.1145/3563659.3563664}
\showURL{%
\tempurl}


\bibitem[Lala et~al\mbox{.}(2017)]%
        {lala2017attentive}
\bibfield{author}{\bibinfo{person}{Divesh Lala}, \bibinfo{person}{Pierrick Milhorat}, \bibinfo{person}{Koji Inoue}, \bibinfo{person}{Masanari Ishida}, \bibinfo{person}{Katsuya Takanashi}, {and} \bibinfo{person}{Tatsuya Kawahara}.} \bibinfo{year}{2017}\natexlab{}.
\newblock \showarticletitle{Attentive listening system with backchanneling, response generation and flexible turn-taking}. In \bibinfo{booktitle}{\emph{Proceedings of the 18th Annual SIGdial Meeting on Discourse and Dialogue}}. \bibinfo{pages}{127--136}.
\newblock


\bibitem[Langley and Sheppeard(1985)]%
        {langley1985visual}
\bibfield{author}{\bibinfo{person}{GB Langley} {and} \bibinfo{person}{H Sheppeard}.} \bibinfo{year}{1985}\natexlab{}.
\newblock \showarticletitle{The visual analogue scale: its use in pain measurement}.
\newblock \bibinfo{journal}{\emph{Rheumatology international}} \bibinfo{volume}{5}, \bibinfo{number}{4} (\bibinfo{year}{1985}), \bibinfo{pages}{145--148}.
\newblock


\bibitem[Lee and Marsella(2006)]%
        {lee2006nonverbal}
\bibfield{author}{\bibinfo{person}{Jina Lee} {and} \bibinfo{person}{Stacy Marsella}.} \bibinfo{year}{2006}\natexlab{}.
\newblock \showarticletitle{Nonverbal behavior generator for embodied conversational agents}. In \bibinfo{booktitle}{\emph{International Workshop on Intelligent Virtual Agents}}. Springer, \bibinfo{pages}{243--255}.
\newblock


\bibitem[Lee et~al\mbox{.}(2024)]%
        {lee2024large}
\bibfield{author}{\bibinfo{person}{Yoon~Kyung Lee}, \bibinfo{person}{Jina Suh}, \bibinfo{person}{Hongli Zhan}, \bibinfo{person}{Junyi~Jessy Li}, {and} \bibinfo{person}{Desmond~C Ong}.} \bibinfo{year}{2024}\natexlab{}.
\newblock \showarticletitle{Large Language Models Produce Responses Perceived to be Empathic}.
\newblock \bibinfo{journal}{\emph{arXiv preprint arXiv:2403.18148}} (\bibinfo{year}{2024}).
\newblock


\bibitem[Lin et~al\mbox{.}(2023)]%
        {lin2023toward}
\bibfield{author}{\bibinfo{person}{Eleanor Lin}, \bibinfo{person}{James Hale}, {and} \bibinfo{person}{Jonathan Gratch}.} \bibinfo{year}{2023}\natexlab{}.
\newblock \showarticletitle{Toward a Better Understanding of the Emotional Dynamics of Negotiation with Large Language Models}. In \bibinfo{booktitle}{\emph{Proceedings of the Twenty-fourth International Symposium on Theory, Algorithmic Foundations, and Protocol Design for Mobile Networks and Mobile Computing}}. \bibinfo{pages}{545--550}.
\newblock


\bibitem[Lorié et~al\mbox{.}(2017)]%
        {Lorie2017}
\bibfield{author}{\bibinfo{person}{Á Lorié}, \bibinfo{person}{D.~A. Reinero}, \bibinfo{person}{M. Phillips}, \bibinfo{person}{L. Zhang}, {and} \bibinfo{person}{H. Riess}.} \bibinfo{year}{2017}\natexlab{}.
\newblock \showarticletitle{Culture and nonverbal expressions of empathy in clinical settings: A systematic review}.
\newblock \bibinfo{journal}{\emph{Patient Educ Couns}} \bibinfo{volume}{100}, \bibinfo{number}{3} (\bibinfo{year}{2017}), \bibinfo{pages}{411--424}.
\newblock


\bibitem[Lugrin et~al\mbox{.}(2022)]%
        {10.1145/3563659}
\bibfield{editor}{\bibinfo{person}{Birgit Lugrin}, \bibinfo{person}{Catherine Pelachaud}, {and} \bibinfo{person}{David Traum}} (Eds.). \bibinfo{year}{2022}\natexlab{}.
\newblock \bibinfo{booktitle}{\emph{The Handbook on Socially Interactive Agents: 20 years of Research on Embodied Conversational Agents, Intelligent Virtual Agents, and Social Robotics Volume 2: Interactivity, Platforms, Application} (\bibinfo{edition}{1} ed.)}. Vol.~\bibinfo{volume}{48}.
\newblock \bibinfo{publisher}{Association for Computing Machinery}, \bibinfo{address}{New York, NY, USA}.
\newblock
\showISBNx{9781450398961}


\bibitem[Mahadevan et~al\mbox{.}(2024)]%
        {mahadevan2024generative}
\bibfield{author}{\bibinfo{person}{Karthik Mahadevan}, \bibinfo{person}{Jonathan Chien}, \bibinfo{person}{Noah Brown}, \bibinfo{person}{Zhuo Xu}, \bibinfo{person}{Carolina Parada}, \bibinfo{person}{Fei Xia}, \bibinfo{person}{Andy Zeng}, \bibinfo{person}{Leila Takayama}, {and} \bibinfo{person}{Dorsa Sadigh}.} \bibinfo{year}{2024}\natexlab{}.
\newblock \showarticletitle{Generative expressive robot behaviors using large language models}.
\newblock \bibinfo{journal}{\emph{arXiv preprint arXiv:2401.14673}} (\bibinfo{year}{2024}).
\newblock


\bibitem[Mayer et~al\mbox{.}(2003)]%
        {mayer2003measuring}
\bibfield{author}{\bibinfo{person}{John~D Mayer}, \bibinfo{person}{Peter Salovey}, \bibinfo{person}{David~R Caruso}, {and} \bibinfo{person}{Gill Sitarenios}.} \bibinfo{year}{2003}\natexlab{}.
\newblock \showarticletitle{Measuring emotional intelligence with the MSCEIT V2. 0.}
\newblock \bibinfo{journal}{\emph{Emotion}} \bibinfo{volume}{3}, \bibinfo{number}{1} (\bibinfo{year}{2003}), \bibinfo{pages}{97}.
\newblock


\bibitem[McKeown et~al\mbox{.}(2011)]%
        {mckeown2011semaine}
\bibfield{author}{\bibinfo{person}{Gary McKeown}, \bibinfo{person}{Michel Valstar}, \bibinfo{person}{Roddy Cowie}, \bibinfo{person}{Maja Pantic}, {and} \bibinfo{person}{Marc Schroder}.} \bibinfo{year}{2011}\natexlab{}.
\newblock \showarticletitle{The semaine database: Annotated multimodal records of emotionally colored conversations between a person and a limited agent}.
\newblock \bibinfo{journal}{\emph{IEEE transactions on affective computing}} \bibinfo{volume}{3}, \bibinfo{number}{1} (\bibinfo{year}{2011}), \bibinfo{pages}{5--17}.
\newblock


\bibitem[Miller and Rollnick(2023)]%
        {Miller2023}
\bibfield{author}{\bibinfo{person}{W Miller} {and} \bibinfo{person}{S Rollnick}.} \bibinfo{year}{2023}\natexlab{}.
\newblock \bibinfo{booktitle}{\emph{Motivational Interviewing: Helping People Change and Grow} (\bibinfo{edition}{fourth} ed.)}.
\newblock \bibinfo{publisher}{Guilford Press}, \bibinfo{address}{New York}.
\newblock


\bibitem[Murali et~al\mbox{.}(2023)]%
        {Murali2023}
\bibfield{author}{\bibinfo{person}{P Murali}, \bibinfo{person}{M Arjmand}, \bibinfo{person}{M Volonte}, \bibinfo{person}{Z Li}, \bibinfo{person}{J Griffith}, \bibinfo{person}{M Paasche-Orlow}, {and} \bibinfo{person}{T Bickmore}.} \bibinfo{year}{2023}\natexlab{}.
\newblock \showarticletitle{Towards Automated Pain Assessment using Embodied Conversational Agents}. In \bibinfo{booktitle}{\emph{The Third International Workshop on Automated Assessment of Pain (AAP)}}.
\newblock


\bibitem[Nakano et~al\mbox{.}(2003)]%
        {nakano2003towards}
\bibfield{author}{\bibinfo{person}{Yukiko~I Nakano}, \bibinfo{person}{Gabe Reinstein}, \bibinfo{person}{Tom Stocky}, {and} \bibinfo{person}{Justine Cassell}.} \bibinfo{year}{2003}\natexlab{}.
\newblock \showarticletitle{Towards a model of face-to-face grounding}. In \bibinfo{booktitle}{\emph{Proceedings of the 41st annual meeting of the Association for Computational Linguistics}}. \bibinfo{pages}{553--561}.
\newblock


\bibitem[Pelachaud et~al\mbox{.}(2021)]%
        {SIABook1}
\bibfield{author}{\bibinfo{person}{C Pelachaud}, \bibinfo{person}{C Busso}, {and} \bibinfo{person}{D Heylen}.} \bibinfo{year}{2021}\natexlab{}.
\newblock \bibinfo{booktitle}{\emph{Multimodal Behavior Modeling for Socially Interactive Agents}}. Vol.~\bibinfo{volume}{1}.
\newblock \bibinfo{publisher}{Association for Computing Machinery}, \bibinfo{pages}{259--310}.
\newblock


\bibitem[Riess et~al\mbox{.}(2012)]%
        {Riess2012}
\bibfield{author}{\bibinfo{person}{H. Riess}, \bibinfo{person}{J.~M. Kelley}, \bibinfo{person}{R.~W. Bailey}, \bibinfo{person}{E.~J. Dunn}, {and} \bibinfo{person}{M. Phillips}.} \bibinfo{year}{2012}\natexlab{}.
\newblock \showarticletitle{Empathy training for resident physicians: a randomized controlled trial of a neuroscience-informed curriculum}.
\newblock \bibinfo{journal}{\emph{J Gen Intern Med}} \bibinfo{volume}{27}, \bibinfo{number}{10} (\bibinfo{year}{2012}), \bibinfo{pages}{1280--6}.
\newblock


\bibitem[Russell and Mehrabian(1977)]%
        {russell1977evidence}
\bibfield{author}{\bibinfo{person}{James~A Russell} {and} \bibinfo{person}{Albert Mehrabian}.} \bibinfo{year}{1977}\natexlab{}.
\newblock \showarticletitle{Evidence for a three-factor theory of emotions}.
\newblock \bibinfo{journal}{\emph{Journal of research in Personality}} \bibinfo{volume}{11}, \bibinfo{number}{3} (\bibinfo{year}{1977}), \bibinfo{pages}{273--294}.
\newblock


\bibitem[Schroder et~al\mbox{.}(2011)]%
        {schroder2011building}
\bibfield{author}{\bibinfo{person}{Marc Schroder}, \bibinfo{person}{Elisabetta Bevacqua}, \bibinfo{person}{Roddy Cowie}, \bibinfo{person}{Florian Eyben}, \bibinfo{person}{Hatice Gunes}, \bibinfo{person}{Dirk Heylen}, \bibinfo{person}{Mark Ter~Maat}, \bibinfo{person}{Gary McKeown}, \bibinfo{person}{Sathish Pammi}, \bibinfo{person}{Maja Pantic}, {et~al\mbox{.}}} \bibinfo{year}{2011}\natexlab{}.
\newblock \showarticletitle{Building autonomous sensitive artificial listeners}.
\newblock \bibinfo{journal}{\emph{IEEE transactions on affective computing}} \bibinfo{volume}{3}, \bibinfo{number}{2} (\bibinfo{year}{2011}), \bibinfo{pages}{165--183}.
\newblock


\bibitem[Shahverdi et~al\mbox{.}(2023)]%
        {Shahverdi2023}
\bibfield{author}{\bibinfo{person}{P Shahverdi}, \bibinfo{person}{K Rousso}, \bibinfo{person}{J Klotz}, \bibinfo{person}{I Bakhoda}, \bibinfo{person}{M Zribi}, {and} \bibinfo{person}{W-Y Louie}.} \bibinfo{year}{2023}\natexlab{}.
\newblock \showarticletitle{Emotionally Specific Backchanneling in Social Human-Robot Interaction and Human-Human Interaction}. In \bibinfo{booktitle}{\emph{IROS}}.
\newblock


\bibitem[Stefanov et~al\mbox{.}(2020)]%
        {10.1145/3382507.3418832}
\bibfield{author}{\bibinfo{person}{Kalin Stefanov}, \bibinfo{person}{Baiyu Huang}, \bibinfo{person}{Zongjian Li}, {and} \bibinfo{person}{Mohammad Soleymani}.} \bibinfo{year}{2020}\natexlab{}.
\newblock \showarticletitle{OpenSense: A Platform for Multimodal Data Acquisition and Behavior Perception}. In \bibinfo{booktitle}{\emph{Proceedings of the 2020 International Conference on Multimodal Interaction}}. \bibinfo{publisher}{Association for Computing Machinery}, \bibinfo{pages}{660–664}.
\newblock


\bibitem[Tak and Gratch(2023)]%
        {tak2023gpt}
\bibfield{author}{\bibinfo{person}{Ala~N Tak} {and} \bibinfo{person}{Jonathan Gratch}.} \bibinfo{year}{2023}\natexlab{}.
\newblock \showarticletitle{Is GPT a Computational Model of Emotion?}. In \bibinfo{booktitle}{\emph{2023 11th International Conference on Affective Computing and Intelligent Interaction (ACII)}}. IEEE, \bibinfo{pages}{1--8}.
\newblock


\bibitem[Touvron et~al\mbox{.}(2023)]%
        {touvron2023llama}
\bibfield{author}{\bibinfo{person}{Hugo Touvron}, \bibinfo{person}{Louis Martin}, \bibinfo{person}{Kevin Stone}, \bibinfo{person}{Peter Albert}, \bibinfo{person}{Amjad Almahairi}, \bibinfo{person}{Yasmine Babaei}, \bibinfo{person}{Nikolay Bashlykov}, \bibinfo{person}{Soumya Batra}, \bibinfo{person}{Prajjwal Bhargava}, \bibinfo{person}{Shruti Bhosale}, {et~al\mbox{.}}} \bibinfo{year}{2023}\natexlab{}.
\newblock \showarticletitle{Llama 2: Open foundation and fine-tuned chat models}.
\newblock \bibinfo{journal}{\emph{arXiv preprint arXiv:2307.09288}} (\bibinfo{year}{2023}).
\newblock


\bibitem[Traum(1994)]%
        {traum1994computational}
\bibfield{author}{\bibinfo{person}{David Traum}.} \bibinfo{year}{1994}\natexlab{}.
\newblock \showarticletitle{A computational theory of grounding in natural language conversation}.
\newblock  (\bibinfo{year}{1994}).
\newblock


\bibitem[Utami and Bickmore(2019)]%
        {utami2019collaborative}
\bibfield{author}{\bibinfo{person}{Dina Utami} {and} \bibinfo{person}{Timothy Bickmore}.} \bibinfo{year}{2019}\natexlab{}.
\newblock \showarticletitle{Collaborative user responses in multiparty interaction with a couples counselor robot}. In \bibinfo{booktitle}{\emph{2019 14th ACM/IEEE International Conference on Human-Robot Interaction (HRI)}}. IEEE, \bibinfo{pages}{294--303}.
\newblock


\bibitem[Wang et~al\mbox{.}(2024)]%
        {wang2024ain}
\bibfield{author}{\bibinfo{person}{Zining Wang}, \bibinfo{person}{Paul Reisert}, \bibinfo{person}{Eric Nichols}, {and} \bibinfo{person}{Randy Gomez}.} \bibinfo{year}{2024}\natexlab{}.
\newblock \showarticletitle{Ain't Misbehavin'--Using LLMs to Generate Expressive Robot Behavior in Conversations with the Tabletop Robot Haru}.
\newblock \bibinfo{journal}{\emph{arXiv preprint arXiv:2402.11571}} (\bibinfo{year}{2024}).
\newblock


\bibitem[Ward(1996)]%
        {ward1996using}
\bibfield{author}{\bibinfo{person}{Nigel Ward}.} \bibinfo{year}{1996}\natexlab{}.
\newblock \showarticletitle{Using prosodic clues to decide when to produce back-channel utterances}. In \bibinfo{booktitle}{\emph{Proceeding of Fourth International Conference on Spoken Language Processing. ICSLP'96}}, Vol.~\bibinfo{volume}{3}. IEEE, \bibinfo{pages}{1728--1731}.
\newblock


\bibitem[Ward and Tsukahara(2000)]%
        {Ward2000}
\bibfield{author}{\bibinfo{person}{N Ward} {and} \bibinfo{person}{W Tsukahara}.} \bibinfo{year}{2000}\natexlab{}.
\newblock \showarticletitle{Prosodic features which cue back-channel responses in English and Japanese}.
\newblock \bibinfo{journal}{\emph{J Pragmatics}} \bibinfo{volume}{32}, \bibinfo{number}{8} (\bibinfo{year}{2000}), \bibinfo{pages}{1177–1207}.
\newblock


\bibitem[Wiener(2019)]%
        {wiener2019cybernetics}
\bibfield{author}{\bibinfo{person}{Norbert Wiener}.} \bibinfo{year}{2019}\natexlab{}.
\newblock \bibinfo{booktitle}{\emph{Cybernetics or Control and Communication in the Animal and the Machine}}.
\newblock \bibinfo{publisher}{MIT press}.
\newblock


\bibitem[Wispe(1986)]%
        {Wispe1986}
\bibfield{author}{\bibinfo{person}{L Wispe}.} \bibinfo{year}{1986}\natexlab{}.
\newblock \showarticletitle{The distinction between sympathy and empathy: To call forth a concept, a word is needed}.
\newblock \bibinfo{journal}{\emph{Journal of personality and social psychology}} \bibinfo{volume}{50}, \bibinfo{number}{2} (\bibinfo{year}{1986}), \bibinfo{pages}{314--321}.
\newblock


\bibitem[Zhan et~al\mbox{.}(2023)]%
        {zhan2023evaluating}
\bibfield{author}{\bibinfo{person}{Hongli Zhan}, \bibinfo{person}{Desmond~C Ong}, {and} \bibinfo{person}{Junyi~Jessy Li}.} \bibinfo{year}{2023}\natexlab{}.
\newblock \showarticletitle{Evaluating subjective cognitive appraisals of emotions from large language models}.
\newblock \bibinfo{journal}{\emph{arXiv preprint arXiv:2310.14389}} (\bibinfo{year}{2023}).
\newblock


\end{thebibliography}



\end{document}